\documentclass[twocolumn]{aastex63}
\usepackage{amsmath}
\usepackage[T1]{fontenc}

\def\logg{{\rm log}\,g}

\newcommand{\kpc}{\ensuremath{\textrm{kpc}}}
\newcommand{\feh}{\ensuremath{\textrm{[Fe/H]}}}
\newcommand{\afe}{\ensuremath{\textrm{[$\alpha$/Fe]}}}

\newcommand{\MS}{\texttt{MINESweeper}}
\newcommand{\gaia}{\textsl{Gaia}}

\newcommand{\hhh}{H3}
\newcommand{\package}[1]{\textsl{#1}}
\newcommand{\paper}{\textsl{Letter}}

\shorttitle{early assembly of the milky way}
\shortauthors{bonaca et al.}

\begin{document}\sloppy\sloppypar\raggedbottom\frenchspacing 
\title{Timing the Early Assembly of the Milky Way with the H3 Survey}

\correspondingauthor{Ana~Bonaca}
\email{ana.bonaca@cfa.harvard.edu}

\author[0000-0002-7846-9787]{Ana~Bonaca}
\affil{Center for Astrophysics | Harvard \& Smithsonian, 60 Garden Street, Cambridge, MA 02138, USA}

\author[0000-0002-1590-8551]{Charlie~Conroy}
\affil{Center for Astrophysics | Harvard \& Smithsonian, 60 Garden Street, Cambridge, MA 02138, USA}

\author[0000-0002-1617-8917]{Phillip~A.~Cargile}
\affil{Center for Astrophysics | Harvard \& Smithsonian, 60 Garden Street, Cambridge, MA 02138, USA}

\author[0000-0003-3997-5705]{Rohan~P.~Naidu}
\affil{Center for Astrophysics | Harvard \& Smithsonian, 60 Garden Street, Cambridge, MA 02138, USA}

\author[0000-0002-9280-7594]{Benjamin~D.~Johnson}
\affil{Center for Astrophysics | Harvard \& Smithsonian, 60 Garden Street, Cambridge, MA 02138, USA}

\author[0000-0002-5177-727X]{Dennis~Zaritsky}
\affil{Steward Observatory and University of Arizona, 933 N. Cherry Ave, Tucson, AZ 85719, USA}

\author[0000-0001-5082-9536]{Yuan-Sen~Ting}
\affil{Institute for Advanced Study, Princeton, NJ 08540, USA}
\affil{Department of Astrophysical Sciences, Princeton University, Princeton, NJ 08544, USA}
\affil{Observatories of the Carnegie Institution of Washington, 813 Santa Barbara Street, Pasadena, CA 91101, USA}
\affil{Research School of Astronomy and Astrophysics, Mount Stromlo Observatory, Cotter Road, Weston Creek, ACT 2611, Canberra, Australia}

\author[0000-0003-2352-3202]{Nelson~Caldwell}
\affil{Center for Astrophysics | Harvard \& Smithsonian, 60 Garden Street, Cambridge, MA 02138, USA}

\author[0000-0002-6800-5778]{Jiwon~Jesse~Han}
\affil{Center for Astrophysics | Harvard \& Smithsonian, 60 Garden Street, Cambridge, MA 02138, USA}

\author[0000-0002-8282-9888]{Pieter~van~Dokkum}
\affil{Astronomy Department, Yale University, 52 Hillhouse Ave, New Haven, CT 06511, USA}

\begin{abstract}\noindent
The archaeological record of stars in the Milky Way opens a uniquely detailed window into the early formation and assembly of galaxies.
Here we use 11,000 main-sequence turn-off stars with well-measured ages, \feh, \afe, and orbits from the H3 Survey and \gaia\ to time the major events in the early Galaxy.
Located beyond the Galactic plane, $1\lesssim |Z|/\kpc \lesssim4$, this sample contains three chemically distinct groups: a low metallicity population, and low-$\alpha$ and high-$\alpha$ groups at higher metallicity.
The age and orbit distributions of these populations show that: 1) the high-$\alpha$ group, which includes both disk stars and the in-situ halo, has a star-formation history independent of eccentricity that abruptly truncated $8.3\pm0.1$\,Gyr ago ($z\simeq1$); 2) the low metallicity population, which we identify as the accreted stellar halo, is on eccentric orbits and its star formation truncated $10.2.^{+0.2}_{-0.1}$\,Gyr ago ($z\simeq2$); 3) the low-$\alpha$ population is primarily on low eccentricity orbits and the bulk of its stars formed less than 8\,Gyr ago.
These results suggest a scenario in which the Milky Way accreted a satellite galaxy at $z\approx2$ that merged with the early disk by $z\approx1$.
This merger truncated star formation in the early high-$\alpha$ disk and perturbed a fraction of that disk onto halo-like orbits.
The merger enabled the formation of a chemically distinct, low-$\alpha$ disk at $z\lesssim1$.
The lack of any stars on halo-like orbits at younger ages indicates that this event was the last significant disturbance to the Milky Way disk.
\vspace{1cm}
\end{abstract}

\section{Introduction}
\label{sec:intro}
Our cosmological paradigm predicts that Milky Way-like galaxies assembled half or more of their stars by $z\sim1$ \citep[e.g.,][]{Moster13, Behroozi13}.
Modern hydrodynamical simulations show that the physical processes driving this evolution, including gas accretion, mergers, and feedback, operate on relatively short timescales and are often manifested on sub-kpc spatial scales \citep[e.g.,][]{Sawala16,Hopkins18,Pillepich19}.
These scales are currently inaccessible to direct observations of galaxies at redshifts $z\gtrsim1$ \citep[e.g.,][]{Nelson16, Tacchella18}.
A promising alternative is to use the archaeological record to gain insight into the formation and evolution of the early Milky Way.

Recent data from the \gaia\ observatory \citep{gdr2} has revealed that the Milky Way likely underwent a significant merger at $z\gtrsim1$ \citep[$\approx$1:10 total mass ratio;][]{Belokurov18, Helmi18, Mackereth19,Kruijssen20}.
Stars accreted in this merger dominate the inner halo of the Galaxy (Naidu et al., in preparation), however, the exact timing of this event and its role in sculpting the Galactic disk is less clear \citep[cf.][]{Gallart19,Belokurov19b,Vincenzo19}.

In this \paper\ we use a large sample of main-sequence turn-off stars from the ``Hectochelle in the Halo at High Resolution'' (H3) Spectroscopic Survey (\S\ref{sec:data}) to identify the main components of the Milky Way (\S\ref{sec:chem}) and precisely measure their star-formation histories (\S\ref{sec:ages}).
In \S\ref{sec:discussion} we use these histories to provide the most detailed accounting of the Milky Way's early assembly.

\begin{figure}[!t]
\begin{center}
\includegraphics[width=0.95\columnwidth]{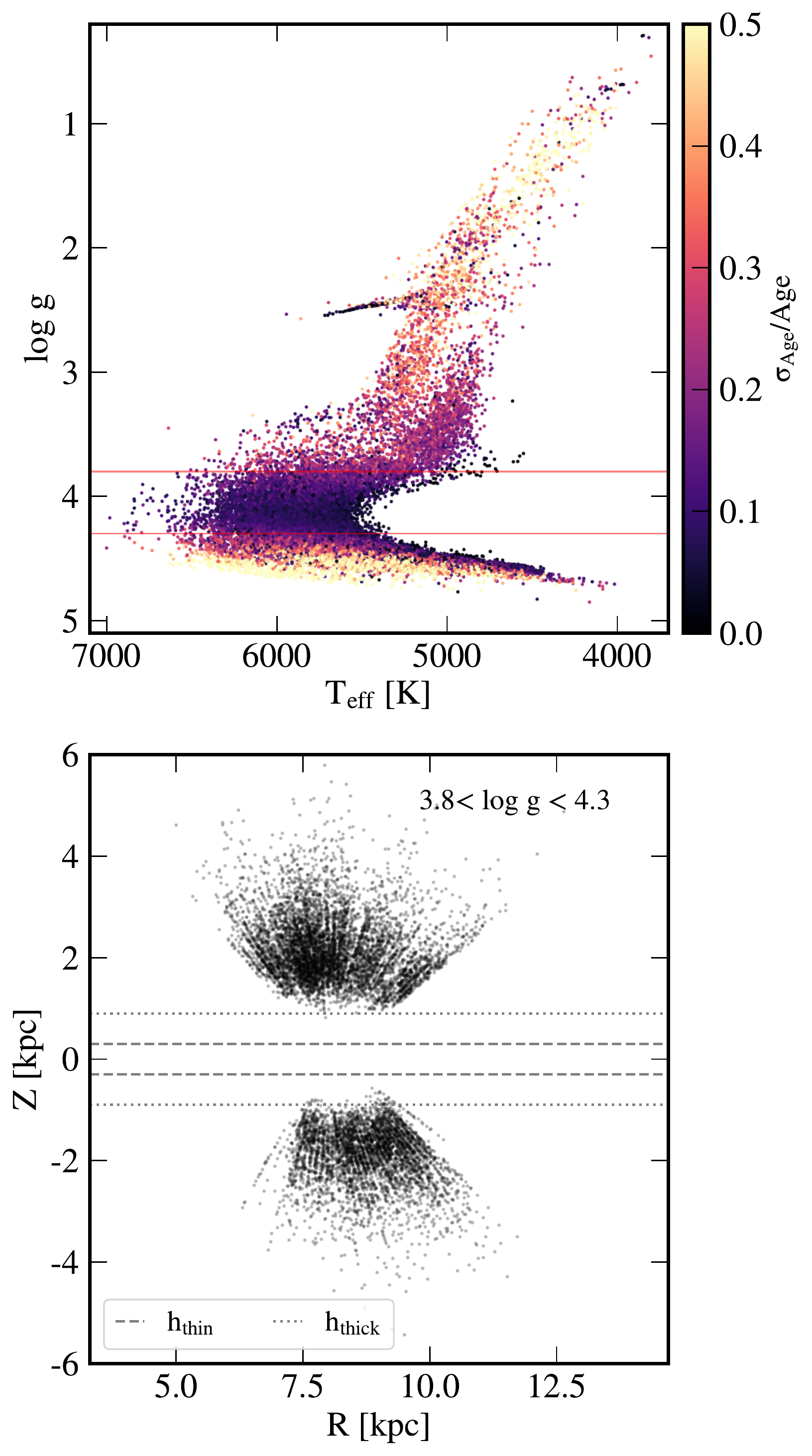}
\caption{
Top panel: Kiel diagram of high signal-to-noise stars from the \hhh\ survey, color-coded by the relative age uncertainty.
Main-sequence turn-off stars, delimited by red horizontal lines ($3.8<\logg<4.3$), have precisely determined ages (the median uncertainty is 10\,\%).
Bottom panel: Cylindrical Galactocentric coordinates of the main-sequence turn-off stars we studied in this work.
Our sample is radially confined to the Solar circle ($6\lesssim R/\kpc \lesssim10$), and vertically extends between $1\lesssim |Z|/\kpc \lesssim4$, beyond the scale heights of the thin and thick disks (dashed and dotted horizontal lines, respectively).
}
\label{fig:logg_selection}
\end{center}
\end{figure}

\section{Data}
\label{sec:data}
We use data from the \hhh\ Survey \citep{Conroy19a}, which is collecting $R\approx32,000$ spectra of distant (\gaia\ parallax $\varpi<0.5$ mas), bright ($15<r<18$) stars at high Galactic latitude ($|b|>40^\circ$) with MMT/Hectochelle \citep{Szentgyorgyi11}.
Stellar parameters including radial velocities, distances, \feh\ and \afe\ abundance ratios, and ages are determined with the Bayesian \MS\ code \citep{Cargile19}.
Briefly, \MS\ fits spectra and photometry using MIST stellar evolution models \citep{Choi16} and stellar spectral models adapted from \texttt{The Payne} framework \citep{Ting19}.
In addition to the directly measured surface abundances, \MS\ reports the inferred \emph{initial} abundances, $\feh_i$ and $\afe_i$.
Here we use the initial abundances, as they are the predicted natal abundances accounting for diffusive and mixing processes during stellar evolution \citep{Dotter17}.
As our main goal is to deduce age differences between different populations of stars in the Milky Way, in this work we computed stellar parameters assuming a uniform age prior between $2-14$ Gyr.
This is in contrast to the fiducial \hhh\ stellar parameter pipeline \citep{Cargile19} and existing frameworks for estimating stellar parameters \citep[e.g.,][]{Das19}, which adopt a Galactic model that specifies (often strong) priors for stellar ages in different components of the Galaxy.

We combined H3 radial velocities and distances with Gaia DR2 proper motions \citep{gdr2} to calculate orbits of stars in our sample.
For orbit integrations we assumed a static Milky Way, with the bulge, disk, and halo components defined in the \package{gala} v1.1 package \citep{gala}, and adopted the \package{astropy} v4.0 default reference frame \citep{astropy} to convert the heliocentric to Galactocentric coordinates.
We used the orbital eccentricity, defined as $e=(r_{apo}-r_{peri})/(r_{apo}+r_{peri})$, where $r_{apo}$, $r_{peri}$ are the apocenter and pericenter, respectively, to dynamically classify orbits.

To date, the H3 survey has observed $125,000$ stars, but in this work we focus on the precisely measured subset of $28,000$ stars with spectral signal-to-noise ratio $S/N>10$.
The top panel of Figure~\ref{fig:logg_selection} shows the inferred temperature and surface gravity for these high $S/N$ stars, color-coded by the relative age uncertainty.
Ages are most precisely measured for stars near the main sequence turn-off (MSTO).
To achieve a precise chronology of the Galaxy, we select $11,034$ MSTO stars with $3.8<\logg<4.3$.
The median and 90th percentile age uncertainties of our adopted sample are 10\% and 16\%, respectively.

The spatial distribution of our sample is shown in cylindrical Galactocentric coordinates in the bottom panel of Figure~\ref{fig:logg_selection}.
The H3 targeting strategy and the high $S/N$ cut result in our sample spanning $1\lesssim |Z|/\kpc \lesssim4$, beyond the scale-heights of the geometric thin and thick disks \citep[dashed and dotted lines, respectively,][]{Juric08}.

\begin{figure}[!t]
\begin{center}
\includegraphics[width=0.95\columnwidth]{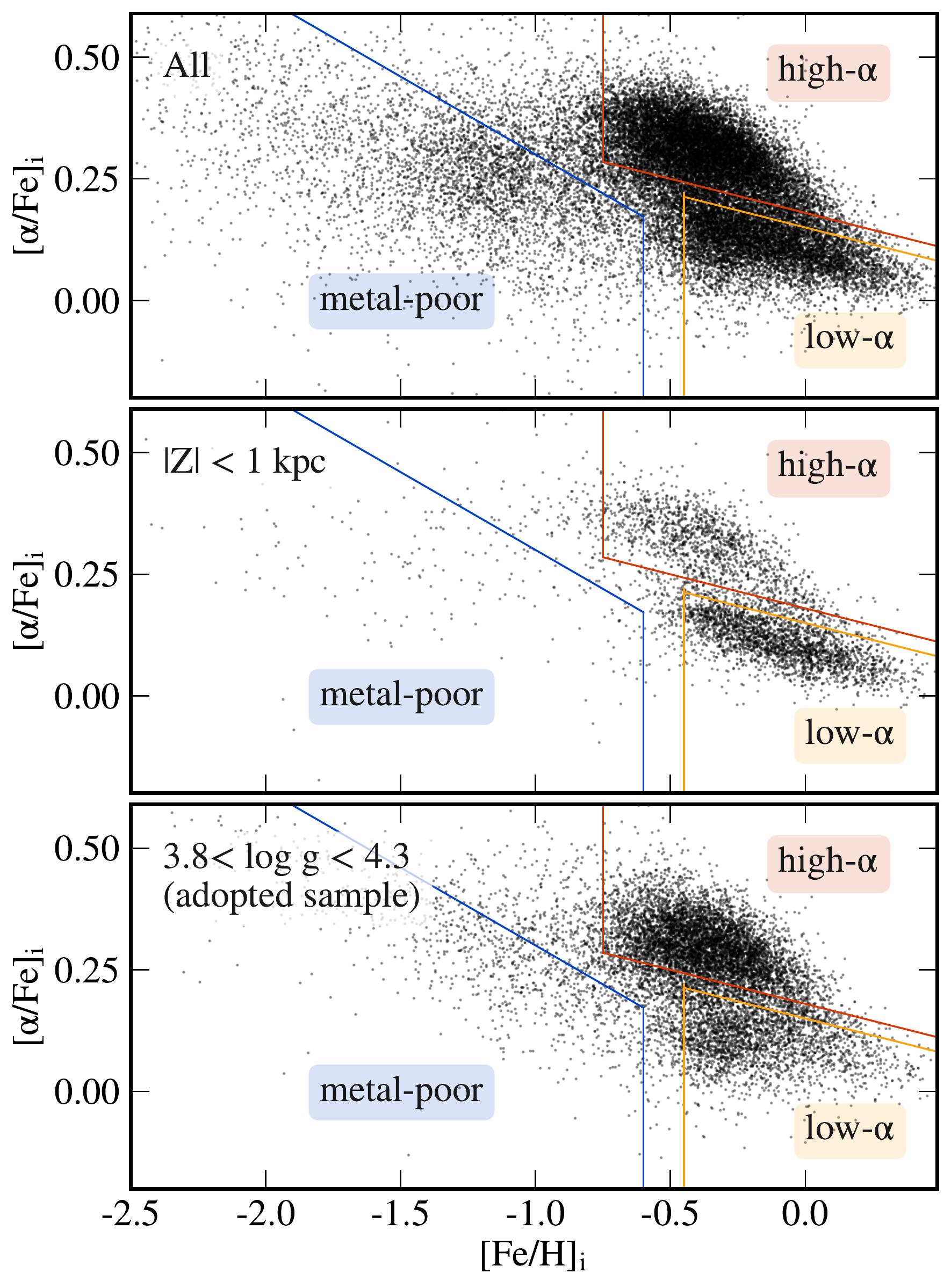}
\caption{
Initial \afe\ vs. initial \feh\ for all high signal-to-noise stars in the H3 survey (top), those within 1\,kpc from the Galactic plane (middle), and our adopted sample of main-sequence turn-off stars (bottom).
All panels clearly show the presence of three distinct components, delineated with colored lines: the \emph{metal-poor} population with a wide range of \afe\ (blue), and the \emph{high-$\alpha$} (red) and \emph{low-$\alpha$} (orange) populations at high metallicities.
}
\label{fig:afeh_selection}
\end{center}
\end{figure}

\begin{figure*}[!t]
\begin{center}
\includegraphics[width=0.9\textwidth]{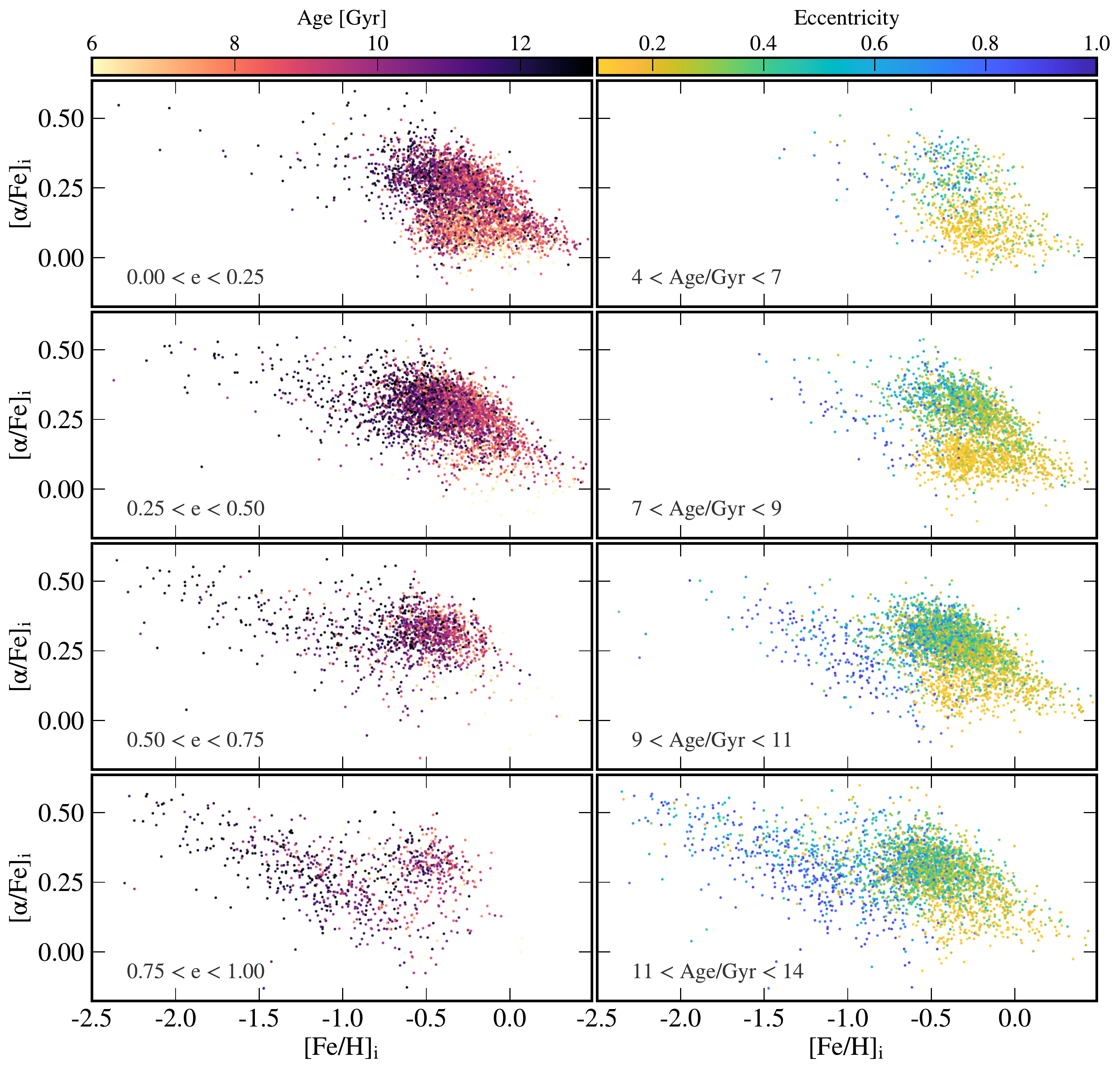}
\caption{
Initial \afe\ vs. initial \feh\ for our sample of main-sequence turn-off stars in bins of orbital eccentricity (left) and age (right).
Chemically distinct components have different orbital properties: metal-poor stars are eccentric, low-$\alpha$ metal-rich stars are circular, while the high-$\alpha$ metal-rich component spans the entire range of eccentricities.
These components also formed at different times: metal-poor stars are ancient (median age of 11.5\,Gyr, and few younger than 9\,Gyr), while the star formation histories of the high- and low-$\alpha$ metal-rich are more extended.
}
\label{fig:afeh_splits}
\end{center}
\end{figure*}

\section{Chemistry and dynamics}
\label{sec:chem}

It has long been known that the Milky Way harbors chemically distinct stellar populations \citep[e.g.,][]{Gilmore89}.
We therefore begin by exploring the chemistry of H3 stars in Figure~\ref{fig:afeh_selection}.
The $\feh_i$ vs. $\afe_i$ abundance ratios for all high $S/N$ stars in H3 (top panel of Figure~\ref{fig:afeh_selection}) reveal three distinct populations: a \emph{metal-poor} population with a wide distribution of \afe\ abundances, and, at high metallicities, distinct \emph{high-$\alpha$} and \emph{low-$\alpha$} populations.
These populations are chemically well-separated, and we define simple selection criteria for each; metal-poor (delineated in blue): $\afe_i<-0.32\times\feh_i-0.02$ and $\feh_i<-0.6$; high-$\alpha$ (red): $\afe_i>-0.14\times\feh_i+0.18$ and $\feh_i>-0.75$; and low-$\alpha$ (orange): $\afe_i<-0.14\times\feh_i+0.15$ and $\feh_i>-0.45$.

The middle panel of Figure~\ref{fig:afeh_selection} shows the abundances of all high $S/N$ stars near the Galactic plane ($|Z|<1$\,kpc).
In this sample, the metal-rich populations are sharply defined and form declining sequences of \afe\ with \feh, in broad agreement with other observations \citep[e.g.,][]{Adibekyan12, Bensby14} and chemical evolution models \citep[e.g.,][]{Kobayashi06}.

The abundance pattern of our MSTO sample is shown in the bottom panel of Figure~\ref{fig:afeh_selection} and features all three of the identified populations.
The low-metallicity and high-$\alpha$ populations have similar patterns to those in the Galactic plane (middle) and in the sample overall (top).
However, the low-$\alpha$ population in MSTO stars has a noticeably larger \afe\ spread than in the plane.
This may be related to the MSTO sample's location above the Galactic plane and will be investigated in the future.

Precise stellar ages enable us to directly trace the chemical and dynamical evolution of the Milky Way.
The left column of Figure~\ref{fig:afeh_splits} shows $\afe_i$ vs. $\feh_i$ colored by age in bins of orbital eccentricity $e$.
High-$\alpha$ metal-rich stars can be found at all eccentricities, in contrast to the low-$\alpha$ population, which is confined to circular orbits ($e\lesssim0.25$), and the metal-poor population, which is found on fairly eccentric orbits ($e\gtrsim0.5$).
In general, metallicity increases with time, but the low-$\alpha$ sequence is younger than the high-$\alpha$ sequence at the same metallicity.
The age distribution of each component is independent of eccentricity, which has implications for the formation of the Galaxy (see \S\,\ref{sec:discussion}).

The right column of Figure~\ref{fig:afeh_splits} shows how these populations evolved in time, with each panel containing a narrow range of ages, color-coded by eccentricity.
Different chemical populations correspond to different distributions of eccentricities, as most clearly highlighted in the bottom two panels.
Comparison of different panels shows that the mapping between the chemistry and eccentricity of a population holds at all ages.

\begin{figure*}[!t]
\begin{center}
\includegraphics[width=0.75\textwidth]{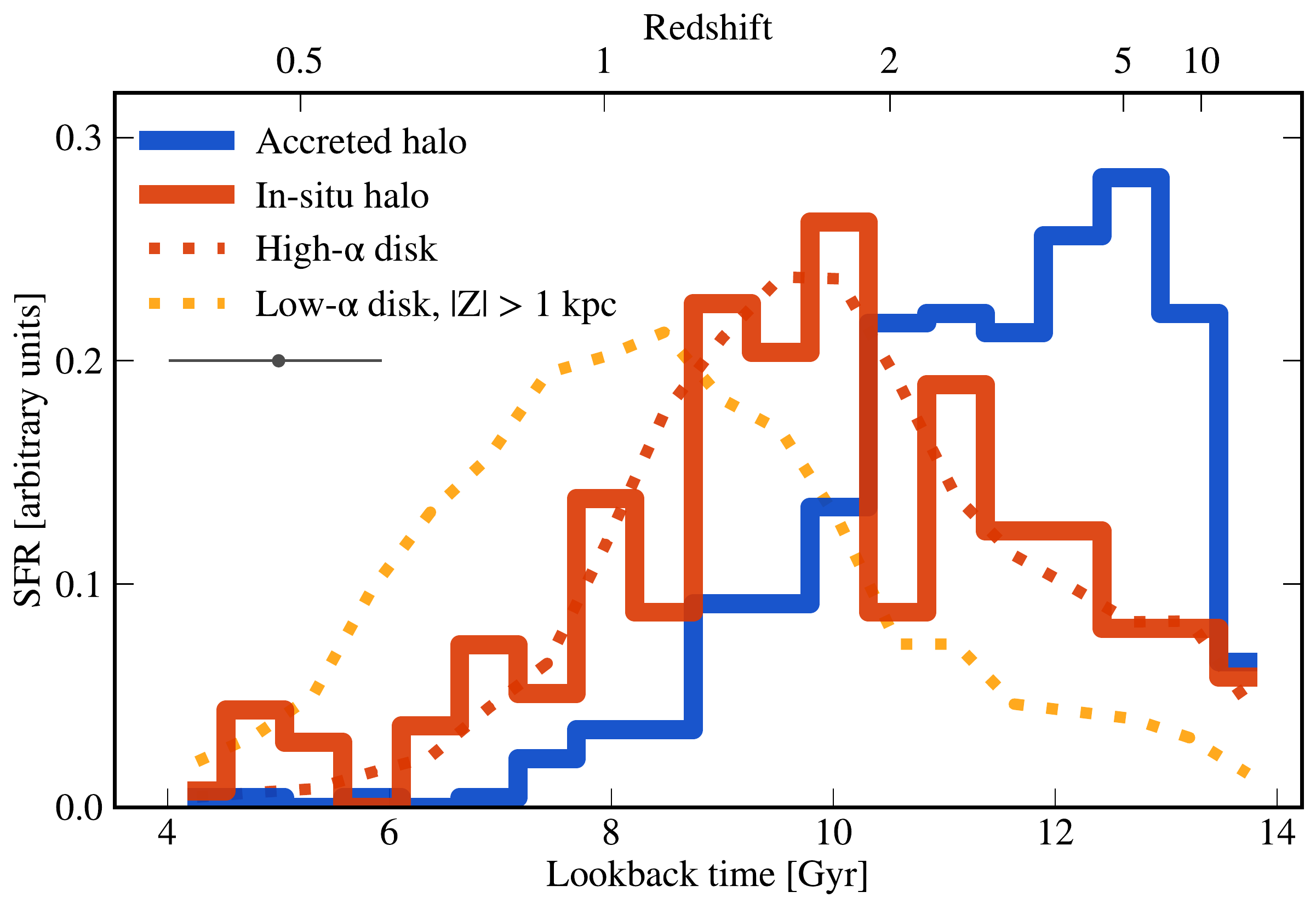}
\caption{
Star-formation histories of chemo-dynamical components in the Milky Way derived from MSTO ages at $1\lesssim|Z|\lesssim4$ kpc.
The accreted halo (metal-poor stars on eccentric orbits; blue line) is older than the in-situ halo (high-$\alpha$ stars on eccentric orbits; red line).
The star-formation history of the in-situ halo is very similar to that of high-$\alpha$ stars on disk-like, circular orbits (red dotted line).
The MSTO stars above the Galactic plane show that the low-$\alpha$ disk forms last (orange dotted line), although they are likely not representative of the global low-$\alpha$ population, which we expect continues forming stars at the present.
The typical age uncertainty for a single star is $\approx1$\,Gyr, shown with an error bar below the legend, so the detected difference in the distribution of ages between the in-situ and the accreted halo is significant.
}
\label{fig:ages}
\end{center}
\end{figure*}

\begin{figure*}
\begin{center}
\includegraphics[width=0.99\textwidth]{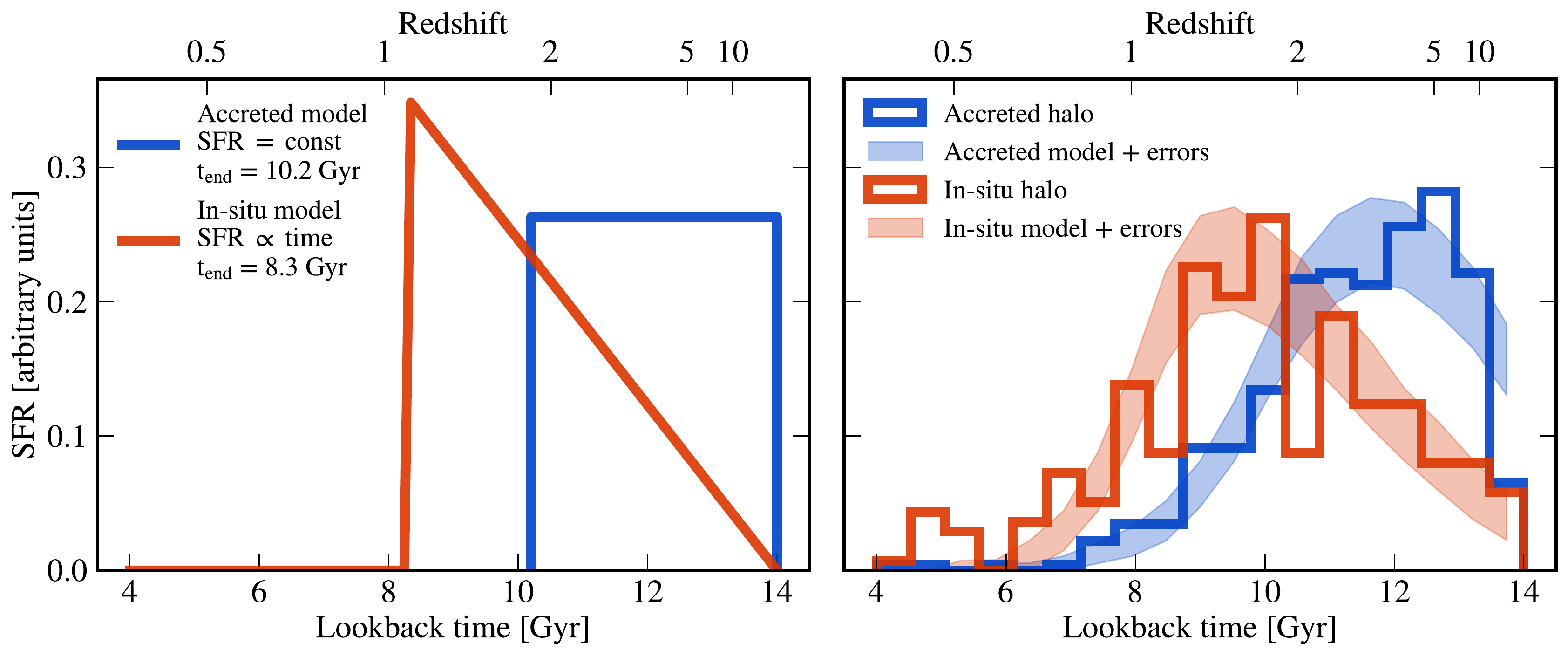}
\caption{
Left panel: Best-fit models for the age distribution of the accreted (blue, constant star-formation rate) and the in-situ halo (red, rising star-formation history).
Right panel: Best-fit star-formation histories convolved with observational uncertainties (shaded) match the inferred distribution of ages for both the in-situ and accreted halo (solid lines).
The inferred ages of local halo stars are consistent with an abrupt end of star formation $\approx10$\,Gyr ago in the accreted halo and $\approx2$\,Gyr later in the in-situ halo.
}
\label{fig:models}
\end{center}
\end{figure*}

\section{Chronology}
\label{sec:ages}

In this section we present star formation histories for several of the Milky Way's structural components.
Motivated by eccentricity patterns in the \afe\ vs. \feh\ space (Figure~\ref{fig:afeh_splits}), and previous chemo-dynamical studies of the Galaxy \citep[e.g.,][]{Hayden15, Bonaca17}, we define: (1) the \emph{low-$\alpha$ disk} as low-$\alpha$ stars with low eccentricity $e<0.25$, (2) the \emph{high-$\alpha$ disk} as high-$\alpha$ stars with low eccentricity $e<0.5$, (3) the \emph{in-situ halo} as high-$\alpha$ stars with high eccentricity $e>0.75$, and (4) the \emph{accreted halo} as metal-poor stars with high eccentricity $e>0.75$.
These selections result in 1818, 4631, 266, and 438 stars in each group, respectively.
We expect these selections, as well as our earlier MSTO sample selection, to be largely unbiased in age, and therefore interpret the distribution of stellar ages in each component as its star formation history.

Figure~\ref{fig:ages} presents the star-formation history beyond the Galactic plane.
Each component is normalized to unity.
The accreted halo (blue line) is ancient, with an approximately constant star-formation rate until 10\,Gyr ago, followed by a sharp decline.
The in-situ halo (red solid line) had a rising SFR that peaked 10\,Gyr ago, after the decline in the accreted halo.
Remarkably, the high-$\alpha$ disk (red dotted line) has a very similar star-formation history to the in-situ halo.
The two-sample K-S statistic for their ages is 0.07---consistent with being drawn from the same distribution.
The low-$\alpha$ disk (orange dotted line) formed last, with the largest increase in star formation coinciding with the drop in star formation of the accreted halo.

Figure~\ref{fig:ages} suggests that the accreted halo is older than the in-situ halo, but to quantify the difference in their ages we need to account for the measurement uncertainties.
We fit the age at which each component stopped forming stars, $t_{end}$, assuming a constant star-formation rate for the accreted halo, and a star-formation rate that rises linearly with time for the in-situ halo.
We chose these one-parameter models for their simplicity.
For a trial $t_{end}$, we drew random samples of ages reflecting the total number of stars in a component, convolved them with the corresponding age uncertainties, and repeated the procedure 50 times.
We compared the median binned distribution of sampled ages to the observed distribution, and constrained $t_{end}$ by sampling the posterior distribution with 64 walkers advanced for 2048 steps using the affine-invariant sampler \package{emcee} \citep{emcee}.
After discarding the first 500 burn-in steps, we report the median $t_{end}$ as the best-fit value, and the $16-84$ percentile range as its uncertainty.
The accreted halo stopped forming stars $10.2^{+0.2}_{-0.1}\,$Gyr ago, while the in-situ halo halted star-formation $8.3\pm0.1\,$Gyr ago.
The best-fit models are shown as solid lines in the left panel of Figure~\ref{fig:models} (blue and red for the accreted and in-situ halo, respectively), and their error convolutions are the shaded regions in the right panel (representing the $16-84$ percentile range).
The best-fit models provide an excellent match to the observed star-formation histories (solid lines, right).
Despite individual age uncertainties of 1\,Gyr, our data show that the in-situ halo stopped forming stars $1.9\pm0.2\,$Gyr after the accreted halo---an age difference detected at $\gtrsim9\,\sigma$.

\section{Discussion}
\label{sec:discussion}

In this work we used data from the H3 Survey and \gaia\ to measure main-sequence turn-off ages for $\approx11,000$ stars located $\gtrsim1\,$kpc beyond the Galactic plane.
We identified the accreted halo as metal-poor stars on eccentric orbits and found they have uniformly old ages, consistent with a sudden truncation $\approx10$\,Gyr ago.
Metal-rich, \afe-enhanced stars orbit with eccentricities ranging from circular to radial, and we associated these extremes with the high-$\alpha$ disk and the in-situ halo, respectively.
These components have the same age distribution, consistent with a linearly rising star-formation rate until $\approx8$\,Gyr ago, followed by a sudden truncation.
The youngest component, that formed the majority of its stars in the last 10\,Gyr, is the low-$\alpha$ disk, defined as metal-rich, low \afe\ stars which are all on close-to-circular orbits.

\begin{figure*}
\begin{center}
\includegraphics[width=0.99\textwidth]{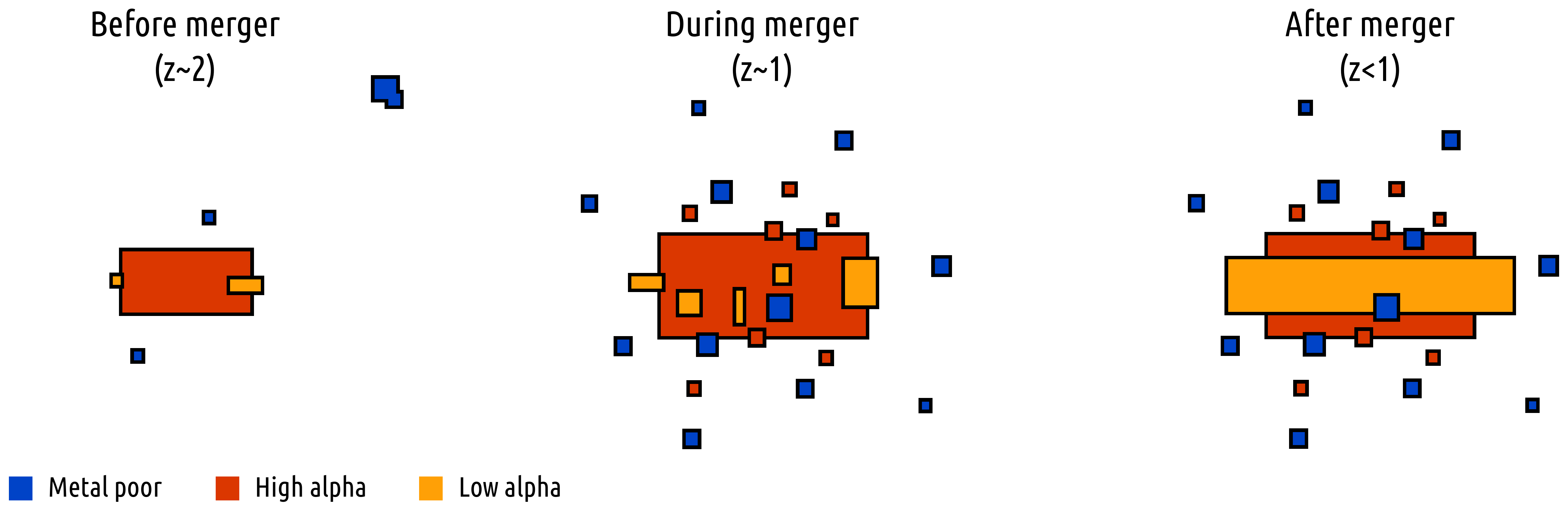}
\caption{
A visual summary of the Milky Way's early assembly.
Left: At early times, the Milky Way was dominated by a hot disk of high-$\alpha$ stars (red), with some metal-poor stars in the halo (blue) and some low-$\alpha$ stars in the disk (orange).
Middle: The merger at $z\approx1$ populated most of the halo with the accreted metal-poor stars, but also heated some in-situ high-$\alpha$ stars to halo-like orbits.
Right: Following the merger, only the low-$\alpha$ disk continues to grow, fueled by gas accreted during the merger.
}
\label{fig:cartoon}
\end{center}
\end{figure*}

These results suggest the following timeline for the early assembly of the Milky Way (Figure~\ref{fig:cartoon}).
During the first $\approx3$\,Gyr, stars were forming in a high-$\alpha$ disk. 
Approximately 10 Gyr ago, a smaller, more metal-poor galaxy was accreted into the Milky Way system.
The satellite lost its gas, which truncated its own star formation, and provided a reservoir of low-metallicity gas that started fueling star formation in a diluted low-$\alpha$ disk.
The satellite orbit gradually decayed, and eventually merged with the Milky Way galaxy 8 Gyr ago.
The merger disrupted star formation in the primordial Milky Way's disk and heated a fraction of its stars to high eccentricity orbits (the in-situ halo).
This scenario is broadly consistent with studies that timed the formation of the Milky Way's low and high-$\alpha$ disks \citep[e.g.,][]{Haywood13, Buder19}, its accreted halo \citep[][]{Belokurov18,Helmi18,Chaplin20}, and those which argued that the in-situ halo stars were kicked out of the disk \citep[e.g.,][]{Zolotov09, Purcell10, Bonaca17, Gallart19, Belokurov19b}.
Below we detail where our results depart from earlier work.

The first key conclusion from this work is that the in-situ halo has the same chemistry and star formation history as the high-$\alpha$ disk.
Previous studies found that the in-situ halo is older than the disk \citep[e.g.,][]{Gallart19, Belokurov19b}.
At least a part of this discrepancy can be traced to the selection of disk stars: simple cuts on height above the Galactic plane \citep{Gallart19} or \feh\ metallicity \citep{Belokurov19b} allow for a significant fraction of younger low-$\alpha$ disk stars.
The age-and-chemistry similarity disfavors models in which the in-situ halo formed kinematically decoupled from the disk, for example, out of gas accreted from infalling satellites \citep[e.g.,][]{Font11, Cooper15}.
The star formation history of the in-situ halo and the high-$\alpha$ disk extends over $\approx5$\,Gyr.
This fact rules out models in which the thick disk formed in a merger-triggered starburst \citep[e.g.,][]{Fantin19}, and limits the fraction of the in-situ halo born in galactic outflows \citep{Yu20}.
Lastly, the sharp decline of star formation disfavors secular scenarios for the formation of the thick disk and the in-situ halo as its high-eccentricity tail \citep[e.g.,][]{Schonrich09, Loebman2011}.

The second key conclusion emerging from the presented data is that the star formation in the in-situ halo ended $\approx2$\,Gyr {\it after} the accreted halo stopped forming stars.
This is in contrast to previous age estimates from \gaia\ photometry that inferred nearly coeval populations \citep{Gallart19}, and from a variety of ground-based spectroscopy that found the in-situ halo to be $\lesssim1$\,Gyr younger than the accreted halo \citep{Belokurov19b}.
These studies interpreted similar ages as evidence that a single event---the last massive merger--- shut star formation in the accreted and the in-situ halo.
Our age estimates, based on a self-consistent analysis of a single spectroscopic dataset, present a more nuanced scenario.
The 2\,Gyr difference between the end of star formation in the in-situ and accreted halo points to different quenching mechanisms.
In our proposed scenario, the satellite galaxy lost its gas shortly after infall, which shut down its own star formation.
The satellite's orbit decayed over 2 Gyr and eventually merged with the Milky Way at $z\approx1$, which quenched in-situ star formation in the high-$\alpha$ disk.
We note that a simple estimate of the dynamical friction timescale for a 1:10 merger at $z=2$ yields an inspiral time of 2 Gyr.
Another possibility is that the satellite stopped forming stars on its own accord 2\,Gyr before merging with the Milky Way and quenching the in-situ star formation.
An independent estimate of the merger timeline \citep[e.g., from properties of accreted globular clusters,][]{Kruijssen20} might distinguish between these scenarios, and probe star formation in a galaxy an order of magnitude lower mass than currently observable at these redshifts \citep[e.g.,][]{Whitaker14}.

Finally, we note the surprising existence of stars older than 10\,Gyr in the low-$\alpha$ disk.
Careful inspection of these stars reveals no issues with the data or derived parameters.
One concern is subtle contamination due to unresolved binaries, which would make the system appear redder and brighter, and bias the inference towards older ages; multi-epoch spectroscopy is required to rule out this possibility.
If confirmed as ancient stars, this would add further complexity to the formation of the early Galaxy.

We close by placing these results in the context of independent constraints on the assembly history of Milky Way-like galaxies.
In this work we have argued that the high-$\alpha$ disk formed by $z\approx1$ (8\,Gyr ago), and subsequent star formation has happened largely in the low-$\alpha$ disk.
\citet{Haywood16} has shown that the mass in the central region of the Milky Way (bulge and/or bar) was also largely assembled by $z\approx1$.  
In a recent review, \citet{Bland-Hawthorn16} give the following masses for these components: $M_{\rm bulge}=1.5\times10^{10}\,M_\odot$, $M_{\rm thick}=6\times10^9\,M_\odot$, and $M_{\rm thin}=3\times10^{10}\,M_\odot$.
Therefore, approximately $40$\% of the Milky Way stellar mass was assembled by $z\approx1$.
As the bulge region harbors some stars younger than 8\,Gyr, and the low-$\alpha$ disk may contain stars older than 8\,Gyr, we treat the 40\% number as tentative without a more detailed accounting of the star-formation history.
Nonetheless, this fraction is in excellent agreement with studies that connect Milky Way-like progenitors across redshift \citep[e.g.,][]{vanDokkum13,Behroozi13,Moster13}.
The Milky Way appears to have grown its mass in a manner that is typical for its mass.
This fact bodes well for making strong connections between the archaeological record of our Galaxy and studies of the high-redshift Universe.

\vspace{0.5cm}
It is a pleasure to thank Kim-Vy Tran and Shy Genel for insightful conversations.
YST is supported by the NASA Hubble Fellowship grant HST-HF2-51425.001 awarded by the Space Telescope Science Institute.
We thank the Hectochelle operators Chun Ly, ShiAnne Kattner, Perry Berlind, and Mike Calkins, and the CfA and U. Arizona TACs for their continued support of the H3 Survey.
Observations reported here were obtained at the MMT Observatory, a joint facility of the Smithsonian Institution and the University of Arizona.
The computations in this \paper\ were run on the FASRC Cannon cluster supported by the FAS Division of Science Research Computing Group at Harvard University.

\facility{MMT(Hectochelle), Gaia}

\software{
\package{Astropy} \citep{astropy, astropy:2018},
\package{colossus} \citep{colossus},
\package{emcee} \citep{emcee},
\package{IPython} \citep{ipython},
\package{matplotlib} \citep{mpl},
\package{numpy} \citep{numpy},
\package{scipy} \citep{scipy}
}

\bibliographystyle{aasjournal}
\bibliography{refs}

\begin{thebibliography}{}
\expandafter\ifx\csname natexlab\endcsname\relax\def\natexlab#1{#1}\fi
\providecommand{\url}[1]{\href{#1}{#1}}

\bibitem[{{Adibekyan} {et~al.}(2012){Adibekyan}, {Sousa}, {Santos}, {Delgado
  Mena}, {Gonz{\'a}lez Hern{\'a}ndez}, {Israelian}, {Mayor}, \&
  {Khachatryan}}]{Adibekyan12}
{Adibekyan}, V.~Z., {Sousa}, S.~G., {Santos}, N.~C., {et~al.} 2012, \aap, 545,
  A32

\bibitem[{{Astropy Collaboration} {et~al.}(2013){Astropy Collaboration},
  {Robitaille}, {Tollerud}, {Greenfield}, {Droettboom}, {Bray}, {Aldcroft},
  {Davis}, {Ginsburg}, {Price-Whelan}, {Kerzendorf}, {Conley}, {Crighton},
  {Barbary}, {Muna}, {Ferguson}, {Grollier}, {Parikh}, {Nair}, {Unther},
  {Deil}, {Woillez}, {Conseil}, {Kramer}, {Turner}, {Singer}, {Fox}, {Weaver},
  {Zabalza}, {Edwards}, {Azalee Bostroem}, {Burke}, {Casey}, {Crawford},
  {Dencheva}, {Ely}, {Jenness}, {Labrie}, {Lim}, {Pierfederici}, {Pontzen},
  {Ptak}, {Refsdal}, {Servillat}, \& {Streicher}}]{astropy}
{Astropy Collaboration}, {Robitaille}, T.~P., {Tollerud}, E.~J., {et~al.} 2013,
  \aap, 558, A33

\bibitem[{{Behroozi} {et~al.}(2013){Behroozi}, {Wechsler}, \&
  {Conroy}}]{Behroozi13}
{Behroozi}, P.~S., {Wechsler}, R.~H., \& {Conroy}, C. 2013, \apj, 770, 57

\bibitem[{{Belokurov} {et~al.}(2018){Belokurov}, {Erkal}, {Evans}, {Koposov},
  \& {Deason}}]{Belokurov18}
{Belokurov}, V., {Erkal}, D., {Evans}, N.~W., {Koposov}, S.~E., \& {Deason},
  A.~J. 2018, \mnras, 478, 611

\bibitem[{{Belokurov} {et~al.}(2019){Belokurov}, {Sanders}, {Fattahi}, {Smith},
  {Deason}, {Evans}, \& {Grand }}]{Belokurov19b}
{Belokurov}, V., {Sanders}, J.~L., {Fattahi}, A., {et~al.} 2019, arXiv
  e-prints, arXiv:1909.04679

\bibitem[{{Bensby} {et~al.}(2014){Bensby}, {Feltzing}, \& {Oey}}]{Bensby14}
{Bensby}, T., {Feltzing}, S., \& {Oey}, M.~S. 2014, \aap, 562, A71

\bibitem[{{Bland-Hawthorn} \& {Gerhard}(2016)}]{Bland-Hawthorn16}
{Bland-Hawthorn}, J., \& {Gerhard}, O. 2016, Annual Review of Astronomy and
  Astrophysics, 54, 529

\bibitem[{{Bonaca} {et~al.}(2017){Bonaca}, {Conroy}, {Wetzel}, {Hopkins}, \&
  {Kere{\v{s}}}}]{Bonaca17}
{Bonaca}, A., {Conroy}, C., {Wetzel}, A., {Hopkins}, P.~F., \& {Kere{\v{s}}},
  D. 2017, \apj, 845, 101

\bibitem[{{Buder} {et~al.}(2019){Buder}, {Lind}, {Ness}, {Asplund}, {Duong},
  {Lin}, {Kos}, {Casagrande}, {Casey}, {Bland-Hawthorn}, {de Silva}, {D'Orazi},
  {Freeman}, {Martell}, {Schlesinger}, {Sharma}, {Simpson}, {Zucker},
  {Zwitter}, {{\v{C}}otar}, {Dotter}, {Hayden}, {Hyde}, {Kafle}, {Lewis},
  {Nataf}, {Nordlander}, {Reid}, {Rix}, {Sk{\'u}lad{\'o}ttir}, {Stello},
  {Ting}, {Traven}, {Wyse}, \& {Galah Collaboration}}]{Buder19}
{Buder}, S., {Lind}, K., {Ness}, M.~K., {et~al.} 2019, \aap, 624, A19

\bibitem[{{Cargile} {et~al.}(2019){Cargile}, {Conroy}, {Johnson}, {Ting},
  {Bonaca}, \& {Dotter}}]{Cargile19}
{Cargile}, P.~A., {Conroy}, C., {Johnson}, B.~D., {et~al.} 2019, arXiv
  e-prints, arXiv:1907.07690

\bibitem[{{Chaplin} {et~al.}(2020){Chaplin}, {Serenelli}, {Miglio}, {Morel},
  {Mackereth}, {Vincenzo}, {Kjeldsen}, {Basu}, {Ball}, {Stokholm}, {Verma},
  {Mosumgaard}, {Silva Aguirre}, {Mazumdar}, {Ranadive}, {Antia}, {Lebreton},
  {Ong}, {Appourchaux}, {Bedding}, {Christensen-Dalsgaard}, {Creevey},
  {Garc{\'\i}a}, {Handberg}, {Huber}, {Kawaler}, {Lund}, {Metcalfe}, {Stassun},
  {Bazot}, {Beck}, {Bell}, {Bergemann}, {Buzasi}, {Benomar}, {Bossini},
  {Bugnet}, {Campante}, {Orhan}, {Corsaro}, {Gonz{\'a}lez-Cuesta}, {Davies},
  {Di Mauro}, {Egeland}, {Elsworth}, {Gaulme}, {Ghasemi}, {Guo}, {Hall},
  {Hasanzadeh}, {Hekker}, {Howe}, {Jenkins}, {Jim{\'e}nez}, {Kiefer},
  {Kuszlewicz}, {Kallinger}, {Latham}, {Lundkvist}, {Mathur}, {Montalb{\'a}n},
  {Mosser}, {Bed{\'o}n}, {Nielsen}, {{\"O}rtel}, {Rendle}, {Ricker},
  {Rodrigues}, {Roxburgh}, {Safari}, {Schofield}, {Seager}, {Smalley},
  {Stello}, {Szab{\'o}}, {Tayar}, {Theme{\ss}l}, {Thomas}, {Vanderspek}, {van
  Rossem}, {Vrard}, {Weiss}, {White}, {Winn}, \& {Y{\i}ld{\i}z}}]{Chaplin20}
{Chaplin}, W.~J., {Serenelli}, A.~M., {Miglio}, A., {et~al.} 2020, Nature
  Astronomy, 7

\bibitem[{{Choi} {et~al.}(2016){Choi}, {Dotter}, {Conroy}, {Cantiello},
  {Paxton}, \& {Johnson}}]{Choi16}
{Choi}, J., {Dotter}, A., {Conroy}, C., {et~al.} 2016, \apj, 823, 102

\bibitem[{{Conroy} {et~al.}(2019){Conroy}, {Bonaca}, {Cargile},
  {et~al.}}]{Conroy19a}
{Conroy}, C., {Bonaca}, A., {Cargile}, P., {et~al.} 2019, \apj, 883, 107

\bibitem[{{Cooper} {et~al.}(2015){Cooper}, {Parry}, {Lowing}, {Cole}, \&
  {Frenk}}]{Cooper15}
{Cooper}, A.~P., {Parry}, O.~H., {Lowing}, B., {Cole}, S., \& {Frenk}, C. 2015,
  \mnras, 454, 3185

\bibitem[{{Das} \& {Sanders}(2019)}]{Das19}
{Das}, P., \& {Sanders}, J.~L. 2019, \mnras, 484, 294

\bibitem[{{Diemer}(2018)}]{colossus}
{Diemer}, B. 2018, \apjs, 239, 35

\bibitem[{{Dotter} {et~al.}(2017){Dotter}, {Conroy}, {Cargile}, \&
  {Asplund}}]{Dotter17}
{Dotter}, A., {Conroy}, C., {Cargile}, P., \& {Asplund}, M. 2017, \apj, 840, 99

\bibitem[{{Fantin} {et~al.}(2019){Fantin}, {C{\^o}t{\'e}}, {McConnachie},
  {Bergeron}, {Cuilland re}, {Gwyn}, {Ibata}, {Thomas}, {Carlberg}, {Fabbro},
  {Haywood}, {Lan{\c{c}}on}, {Lewis}, {Malhan}, {Martin}, {Navarro}, {Scott},
  \& {Starkenburg}}]{Fantin19}
{Fantin}, N.~J., {C{\^o}t{\'e}}, P., {McConnachie}, A.~W., {et~al.} 2019, \apj,
  887, 148

\bibitem[{{Font} {et~al.}(2011){Font}, {McCarthy}, {Crain}, {Theuns}, {Schaye},
  {Wiersma}, \& {Dalla Vecchia}}]{Font11}
{Font}, A.~S., {McCarthy}, I.~G., {Crain}, R.~A., {et~al.} 2011, \mnras, 416,
  2802

\bibitem[{{Foreman-Mackey} {et~al.}(2013){Foreman-Mackey}, {Hogg}, {Lang}, \&
  {Goodman}}]{emcee}
{Foreman-Mackey}, D., {Hogg}, D.~W., {Lang}, D., \& {Goodman}, J. 2013, \pasp,
  125, 306

\bibitem[{{Gaia Collaboration} {et~al.}(2018){Gaia Collaboration}, {Brown},
  {Vallenari}, {Prusti}, {de Bruijne}, {Babusiaux}, {Bailer-Jones}, {Biermann},
  {Evans}, {Eyer}, \& et~al.}]{gdr2}
{Gaia Collaboration}, {Brown}, A.~G.~A., {Vallenari}, A., {et~al.} 2018, \aap,
  616, A1

\bibitem[{{Gallart} {et~al.}(2019){Gallart}, {Bernard}, {Brook}, {Ruiz-Lara},
  {Cassisi}, {Hill}, \& {Monelli}}]{Gallart19}
{Gallart}, C., {Bernard}, E.~J., {Brook}, C.~B., {et~al.} 2019, Nature
  Astronomy, 3, 932

\bibitem[{{Gilmore} {et~al.}(1989){Gilmore}, {Wyse}, \& {Kuijken}}]{Gilmore89}
{Gilmore}, G., {Wyse}, R. F.~G., \& {Kuijken}, K. 1989, \araa, 27, 555

\bibitem[{{Hayden} {et~al.}(2015){Hayden}, {Bovy}, {Holtzman}, {Nidever},
  {Bird}, {Weinberg}, {Andrews}, {Majewski}, {Allende Prieto}, {Anders},
  {Beers}, {Bizyaev}, {Chiappini}, {Cunha}, {Frinchaboy},
  {Garc{\'\i}a-Her{\'n}and ez}, {Garc{\'\i}a P{\'e}rez}, {Girardi}, {Harding},
  {Hearty}, {Johnson}, {M{\'e}sz{\'a}ros}, {Minchev}, {O'Connell}, {Pan},
  {Robin}, {Schiavon}, {Schneider}, {Schultheis}, {Shetrone}, {Skrutskie},
  {Steinmetz}, {Smith}, {Wilson}, {Zamora}, \& {Zasowski}}]{Hayden15}
{Hayden}, M.~R., {Bovy}, J., {Holtzman}, J.~A., {et~al.} 2015, \apj, 808, 132

\bibitem[{{Haywood} {et~al.}(2013){Haywood}, {Di Matteo}, {Lehnert}, {Katz}, \&
  {G{\'o}mez}}]{Haywood13}
{Haywood}, M., {Di Matteo}, P., {Lehnert}, M.~D., {Katz}, D., \& {G{\'o}mez},
  A. 2013, \aap, 560, A109

\bibitem[{{Haywood} {et~al.}(2016){Haywood}, {Di Matteo}, {Snaith}, \&
  {Calamida}}]{Haywood16}
{Haywood}, M., {Di Matteo}, P., {Snaith}, O., \& {Calamida}, A. 2016, \aap,
  593, A82

\bibitem[{{Helmi} {et~al.}(2018){Helmi}, {Babusiaux}, {Koppelman}, {Massari},
  {Veljanoski}, \& {Brown}}]{Helmi18}
{Helmi}, A., {Babusiaux}, C., {Koppelman}, H.~H., {et~al.} 2018, \nat, 563, 85

\bibitem[{{Hopkins} {et~al.}(2018){Hopkins}, {Wetzel}, {Kere{\v{s}}},
  {Faucher-Gigu{\`e}re}, {Quataert}, {Boylan-Kolchin}, {Murray}, {Hayward},
  {Garrison-Kimmel}, {Hummels}, {Feldmann}, {Torrey}, {Ma},
  {Angl{\'e}s-Alc{\'a}zar}, {Su}, {Orr}, {Schmitz}, {Escala}, {Sanderson},
  {Grudi{\'c}}, {Hafen}, {Kim}, {Fitts}, {Bullock}, {Wheeler}, {Chan},
  {Elbert}, \& {Narayanan}}]{Hopkins18}
{Hopkins}, P.~F., {Wetzel}, A., {Kere{\v{s}}}, D., {et~al.} 2018, \mnras, 480,
  800

\bibitem[{{Hunter}(2007)}]{mpl}
{Hunter}, J.~D. 2007, Computing in Science and Engineering, 9, 90

\bibitem[{Jones {et~al.}(2001--)Jones, Oliphant, Peterson, {et~al.}}]{scipy}
Jones, E., Oliphant, T., Peterson, P., {et~al.} 2001--, {SciPy}: Open source
  scientific tools for {Python}, , .
\newblock \url{http://www.scipy.org/}

\bibitem[{{Juri{\'c}} {et~al.}(2008){Juri{\'c}}, {Ivezi{\'c}}, {Brooks},
  {Lupton}, {Schlegel}, {Finkbeiner}, {Padmanabhan}, {Bond}, {Sesar},
  {Rockosi}, {Knapp}, {Gunn}, {Sumi}, {Schneider}, {Barentine}, {Brewington},
  {Brinkmann}, {Fukugita}, {Harvanek}, {Kleinman}, {Krzesinski}, {Long},
  {Neilsen}, {Nitta}, {Snedden}, \& {York}}]{Juric08}
{Juri{\'c}}, M., {Ivezi{\'c}}, {\v{Z}}., {Brooks}, A., {et~al.} 2008, \apj,
  673, 864

\bibitem[{{Kobayashi} {et~al.}(2006){Kobayashi}, {Umeda}, {Nomoto}, {Tominaga},
  \& {Ohkubo}}]{Kobayashi06}
{Kobayashi}, C., {Umeda}, H., {Nomoto}, K., {Tominaga}, N., \& {Ohkubo}, T.
  2006, \apj, 653, 1145

\bibitem[{{Kruijssen} {et~al.}(2020){Kruijssen}, {Pfeffer}, {Chevance},
  {Bonaca}, {Trujillo-Gomez}, {Bastian}, {Reina-Campos}, {Crain}, \&
  {Hughes}}]{Kruijssen20}
{Kruijssen}, J.~M.~D., {Pfeffer}, J.~L., {Chevance}, M., {et~al.} 2020, arXiv
  e-prints, arXiv:2003.01119

\bibitem[{{Loebman} {et~al.}(2011){Loebman}, {Ro{\v{s}}kar}, {Debattista},
  {Ivezi{\'c}}, {Quinn}, \& {Wadsley}}]{Loebman2011}
{Loebman}, S.~R., {Ro{\v{s}}kar}, R., {Debattista}, V.~P., {et~al.} 2011, \apj,
  737, 8

\bibitem[{{Mackereth} {et~al.}(2019){Mackereth}, {Schiavon}, {Pfeffer},
  {et~al.}}]{Mackereth19}
{Mackereth}, J.~T., {Schiavon}, R.~P., {Pfeffer}, J., {et~al.} 2019, \mnras,
  482, 3426

\bibitem[{{Moster} {et~al.}(2013){Moster}, {Naab}, \& {White}}]{Moster13}
{Moster}, B.~P., {Naab}, T., \& {White}, S. D.~M. 2013, \mnras, 428, 3121

\bibitem[{{Nelson} {et~al.}(2016){Nelson}, {van Dokkum}, {F{\"o}rster
  Schreiber}, {Franx}, {Brammer}, {Momcheva}, {Wuyts}, {Whitaker}, {Skelton},
  {Fumagalli}, {Hayward}, {Kriek}, {Labb{\'e}}, {Leja}, {Rix}, {Tacconi}, {van
  der Wel}, {van den Bosch}, {Oesch}, {Dickey}, \& {Ulf Lange}}]{Nelson16}
{Nelson}, E.~J., {van Dokkum}, P.~G., {F{\"o}rster Schreiber}, N.~M., {et~al.}
  2016, \apj, 828, 27

\bibitem[{P\'erez \& Granger(2007)}]{ipython}
P\'erez, F., \& Granger, B.~E. 2007, Computing in Science and Engineering, 9,
  21.
\newblock \url{http://ipython.org}

\bibitem[{{Pillepich} {et~al.}(2019){Pillepich}, {Nelson}, {Springel},
  {Pakmor}, {Torrey}, {Weinberger}, {Vogelsberger}, {Marinacci}, {Genel}, {van
  der Wel}, \& {Hernquist}}]{Pillepich19}
{Pillepich}, A., {Nelson}, D., {Springel}, V., {et~al.} 2019, \mnras, 490, 3196

\bibitem[{{Price-Whelan}(2017)}]{gala}
{Price-Whelan}, A.~M. 2017, The Journal of Open Source Software, 2, 388

\bibitem[{{Price-Whelan} {et~al.}(2018){Price-Whelan}, {Sip{\'{o}}cz},
  {G{\"u}nther}, {Lim}, {Crawford}, {Conseil}, {Shupe}, {Craig}, {Dencheva},
  {Ginsburg}, {VanderPlas}, {Bradley}, {P{\'e}rez-Su{\'a}rez}, {de Val-Borro},
  {Paper Contributors}, {Aldcroft}, {Cruz}, {Robitaille}, {Tollerud},
  {Coordination Committee}, {Ardelean}, {Babej}, {Bach}, {Bachetti}, {Bakanov},
  {Bamford}, {Barentsen}, {Barmby}, {Baumbach}, {Berry}, {Biscani}, {Boquien},
  {Bostroem}, {Bouma}, {Brammer}, {Bray}, {Breytenbach}, {Buddelmeijer},
  {Burke}, {Calderone}, {Cano Rodr{\'\i}guez}, {Cara}, {Cardoso}, {Cheedella},
  {Copin}, {Corrales}, {Crichton}, {D{\textquoteright}Avella}, {Deil},
  {Depagne}, {Dietrich}, {Donath}, {Droettboom}, {Earl}, {Erben}, {Fabbro},
  {Ferreira}, {Finethy}, {Fox}, {Garrison}, {Gibbons}, {Goldstein}, {Gommers},
  {Greco}, {Greenfield}, {Groener}, {Grollier}, {Hagen}, {Hirst}, {Homeier},
  {Horton}, {Hosseinzadeh}, {Hu}, {Hunkeler}, {Ivezi{\'c}}, {Jain}, {Jenness},
  {Kanarek}, {Kendrew}, {Kern}, {Kerzendorf}, {Khvalko}, {King}, {Kirkby},
  {Kulkarni}, {Kumar}, {Lee}, {Lenz}, {Littlefair}, {Ma}, {Macleod},
  {Mastropietro}, {McCully}, {Montagnac}, {Morris}, {Mueller}, {Mumford},
  {Muna}, {Murphy}, {Nelson}, {Nguyen}, {Ninan}, {N{\"o}the}, {Ogaz}, {Oh},
  {Parejko}, {Parley}, {Pascual}, {Patil}, {Patil}, {Plunkett}, {Prochaska},
  {Rastogi}, {Reddy Janga}, {Sabater}, {Sakurikar}, {Seifert}, {Sherbert},
  {Sherwood-Taylor}, {Shih}, {Sick}, {Silbiger}, {Singanamalla}, {Singer},
  {Sladen}, {Sooley}, {Sornarajah}, {Streicher}, {Teuben}, {Thomas},
  {Tremblay}, {Turner}, {Terr{\'o}n}, {van Kerkwijk}, {de la Vega}, {Watkins},
  {Weaver}, {Whitmore}, {Woillez}, {Zabalza}, \& {Contributors}}]{astropy:2018}
{Price-Whelan}, A.~M., {Sip{\'{o}}cz}, B.~M., {G{\"u}nther}, H.~M., {et~al.}
  2018, \aj, 156, 123

\bibitem[{{Purcell} {et~al.}(2010){Purcell}, {Bullock}, \&
  {Kazantzidis}}]{Purcell10}
{Purcell}, C.~W., {Bullock}, J.~S., \& {Kazantzidis}, S. 2010, \mnras, 404,
  1711

\bibitem[{{Sawala} {et~al.}(2016){Sawala}, {Frenk}, {Fattahi}, {Navarro},
  {Bower}, {Crain}, {Dalla Vecchia}, {Furlong}, {Helly}, {Jenkins}, {Oman},
  {Schaller}, {Schaye}, {Theuns}, {Trayford}, \& {White}}]{Sawala16}
{Sawala}, T., {Frenk}, C.~S., {Fattahi}, A., {et~al.} 2016, \mnras, 457, 1931

\bibitem[{{Sch{\"o}nrich} \& {Binney}(2009)}]{Schonrich09}
{Sch{\"o}nrich}, R., \& {Binney}, J. 2009, \mnras, 399, 1145

\bibitem[{{Szentgyorgyi} {et~al.}(2011){Szentgyorgyi}, {Furesz}, {Cheimets},
  {et~al.}}]{Szentgyorgyi11}
{Szentgyorgyi}, A., {Furesz}, G., {Cheimets}, P., {et~al.} 2011, \pasp, 123,
  1188

\bibitem[{{Tacchella} {et~al.}(2018){Tacchella}, {Carollo}, {F{\"o}rster
  Schreiber}, {Renzini}, {Dekel}, {Genzel}, {Lang}, {Lilly}, {Mancini},
  {Onodera}, {Tacconi}, {Wuyts}, \& {Zamorani}}]{Tacchella18}
{Tacchella}, S., {Carollo}, C.~M., {F{\"o}rster Schreiber}, N.~M., {et~al.}
  2018, \apj, 859, 56

\bibitem[{{Ting} {et~al.}(2019){Ting}, {Conroy}, {Rix}, \& {Cargile}}]{Ting19}
{Ting}, Y.-S., {Conroy}, C., {Rix}, H.-W., \& {Cargile}, P. 2019, \apj, 879, 69

\bibitem[{{van Dokkum} {et~al.}(2013){van Dokkum}, {Leja}, {Nelson}, {Patel},
  {Skelton}, {Momcheva}, {Brammer}, {Whitaker}, {Lundgren}, {Fumagalli},
  {Conroy}, {F{\"o}rster Schreiber}, {Franx}, {Kriek}, {Labb{\'e}},
  {Marchesini}, {Rix}, {van der Wel}, \& {Wuyts}}]{vanDokkum13}
{van Dokkum}, P.~G., {Leja}, J., {Nelson}, E.~J., {et~al.} 2013, \apjl, 771,
  L35

\bibitem[{{Vincenzo} {et~al.}(2019){Vincenzo}, {Spitoni}, {Calura},
  {Matteucci}, {Silva Aguirre}, {Miglio}, \& {Cescutti}}]{Vincenzo19}
{Vincenzo}, F., {Spitoni}, E., {Calura}, F., {et~al.} 2019, \mnras, 487, L47

\bibitem[{Walt {et~al.}(2011)Walt, Colbert, \& Varoquaux}]{numpy}
Walt, S. v.~d., Colbert, S.~C., \& Varoquaux, G. 2011, Computing in Science and
  Engg., 13, 22.
\newblock \url{http://dx.doi.org/10.1109/MCSE.2011.37}

\bibitem[{{Whitaker} {et~al.}(2014){Whitaker}, {Franx}, {Leja}, {van Dokkum},
  {Henry}, {Skelton}, {Fumagalli}, {Momcheva}, {Brammer}, {Labb{\'e}},
  {Nelson}, \& {Rigby}}]{Whitaker14}
{Whitaker}, K.~E., {Franx}, M., {Leja}, J., {et~al.} 2014, \apj, 795, 104

\bibitem[{{Yu} {et~al.}(2020){Yu}, {Bullock}, {Wetzel}, {Sand erson}, {Graus},
  {Boylan-Kolchin}, {Nierenberg}, {Grudi{\'c}}, {Hopkins}, {Kere{\v{s}}}, \&
  {Faucher-Gigu{\`e}re}}]{Yu20}
{Yu}, S., {Bullock}, J.~S., {Wetzel}, A., {et~al.} 2020, \mnras,
  arXiv:1912.03316

\bibitem[{{Zolotov} {et~al.}(2009){Zolotov}, {Willman}, {Brooks},
  {et~al.}}]{Zolotov09}
{Zolotov}, A., {Willman}, B., {Brooks}, A.~M., {et~al.} 2009, \apj, 702, 1058

\end{thebibliography}

\end{document}